# Fundamental limitations in spontaneous emission rate of single-photon sources

SERGEY I. BOZHEVOLNYI1,* AND JACOB B. KHURGIN2

1Centre for Nano Optics, University of Southern Denmark, Campusvej 55, DK-5230 Odense M, Denmark

2Johns Hopkins University, Baltimore, Maryland 21218, USA

*Corresponding author: seib@iti.sdu.dk

**Abstract:** Rate of single-photon generation by quantum emitters (QEs) can be enhanced by placing a QE inside a resonant structure. This structure can represent an all-dielectric micro-resonator or waveguide and thus be characterized by ultra-low loss and dimensions on the order of wavelength. Or it can be a metal nanostructure supporting localized or propagating surface plasmon-polariton modes that are of subwavelength dimensions, but suffer from strong absorption. In this work, we develop a physically transparent analytical model of single-photon emission in resonant structures and show unambiguously that, notwithstanding the inherently high loss, the external emission rate can be enhanced with plasmonic nanostructures by two orders of magnitude compared to all-dielectric structures. Our analysis provides guidelines for developments of new plasmonic configurations and materials to be exploited in quantum plasmonics.

Efficient and bright single-photon sources, which enable the generation of single photons with high repetition rates, are crucial components for quantum communication and computation systems [1,2]. The common approach to realization of single-photon sources is to make use of spontaneous emission (SE) from two-level systems emitting one photon at a time – so called quantum emitters (QEs) that can be selected from various atomic or molecular structures, such as dye molecules, quantum dots, color centers in crystals etc.

The intrinsic radiative lifetime $\tau$ of a QE placed within an unconstrained dielectric is of the order of 10 ns in the visible or near IR spectral range, which is certainly too long to ensure high repetition rates of single-photon emission. The SE rate can however be increased by placing a QE in a suitable photonic environment with an increased electromagnetic local density of states [3]. Thus, for a non-absorbing cavity characterized by the quality factor $Q$ and volume $V$ and containing a properly located and oriented QE and being in resonance with the QE radiative transition at the wavelength $\lambda$, the ratio between the modified $\gamma_{SE}$ and free-space $\gamma_0 = 1/\tau$ SE rates, the Purcell factor $F$, is given by [4]

$$F = \frac{\gamma_{SE}}{\gamma_0} = \frac{6}{\pi^2}\left(\frac{\lambda}{2n}\right)^3 \frac{Q}{V} \ , \qquad (1)$$

where $n$ is the medium refractive index inside the cavity.

It is clear from Eq. (1) that the SE rate can be enhanced by using either an optical cavity having a small volume and high finesse or, preferably, both. In recent years, following intensive investigations (see a recent review [2]) two classes of nanostructures have emerged as the candidates for use in SE control and enhancement. The first class is all-dielectric micro-cavities [Fig. 1(a)], including those formed by photonic crystals, in which extremely high quality factors ($Q \geq 10^4$) can be achieved, while the volume remains relatively large, on the order of $(\lambda/2n)^3$ [5,6]. The second class includes "plasmonic nano-cavities" incorporating metals [Fig. 1(b)], in which volumes are much smaller than $(\lambda/2n)^3$ [7], but the Q-factor is typically small ($Q \leq 100$) due to large loss in metals [8]. Despite large volume of work, it is still not clear which route (all-dielectric or plasmonic) can lead to the highest SE rate enhancement. Nor is it apparent whether fundamental limits of SE modification can be found in these configurations. The goal of this Letter is to provide the answers to these questions.

To do so we consider theoretically QE coupling to localized surface plasmons (LSPs) and dielectric micro-cavities from the viewpoint of assessing fundamental factors limiting the achievable SE rates in these configurations. We then compare the QE coupling to propagating surface plasmon-polariton (SPP) modes and to dielectric waveguide modes, arguing that the usage of plasmonic configurations is advantageous in both cases.

For dielectric, i.e., diffraction limited and lossless, cavities one can obtain [Eq. (1)] the following upper limit for the Purcell factor:

$$V \geq \left(\frac{\lambda}{2n}\right)^3 \Rightarrow F \leq \frac{6}{\pi^2} Q \ . \quad (2)$$

The *fundamental* issue with this configuration is related to the fact that an increase in $Q$ is strictly connected with the corresponding decrease in the cavity emission rate: $\gamma_{cav} = \omega/Q$, where $\omega$ is the cavity resonant frequency. This decrease will *inevitably* limit an increase in the SE rate *out of the cavity*, since it cannot exceed the cavity emission rate. Intuitively, the optimum coupling is achieved when the two rates are equal, $\gamma_{SE} \sim \gamma_{cav}$, i.e., each photon emitted into the cavity leaves it before the next one appears and no "bottleneck" is formed. This condition also happens to define the boundary when the QE-cavity coupling enters the strong-coupling regime with energy oscillating coherently between the QE and the cavity in the process of Rabi oscillations [2]. Using the above condition, one can evaluate the optimum quality factor $Q_{opt}$ ensuring the highest out-of-cavity SE rate and, consequently, the *fundamental* limit for the SE emission rate enhanced by a dielectric cavity (see Supplementary Material):

$$Q_{opt} \geq \pi \sqrt{\frac{\omega}{6\gamma_0}}, \text{ and } \gamma_{SE}^{max} \leq \frac{\sqrt{6\omega\gamma_0}}{\pi}. \quad (3)$$

The above condition was obtained by considering the SE modification as described with the Purcell factor [Eq. (1)], which is valid only in the weak-coupling approximation, when $\gamma_{SE} \ll \gamma_{cav}$. It is instructive to show that a similar relation can be found using a more rigorous approach that considers vacuum Rabi oscillations. Introducing the cavity emission into coupled equations describing vacuum Rabi oscillations, one can arrive at the following expression for the out-of-cavity emission rate (see Supplementary Material):

$$r(t) = \frac{4\gamma_{cav}}{\Omega^2} \cdot \left|\frac{s_1 s_2}{s_2 - s_1}\left[e^{s_1 t} - e^{s_2 t}\right]\right|^2 . \quad (4)$$

where $\Omega = \sqrt{6\gamma_0\omega}/\pi$ is the vacuum Rabi frequency in the diffraction-limited cavity, i.e., with the volume $V = (\lambda/2n)^3$, noting that the vacuum Rabi frequency is exactly equal to the maximum SE rate in Eq. (3), and

$$s_{1,2} = -\frac{\gamma_{cav}}{4} \mp \sqrt{\left(\frac{\gamma_{cav}}{4}\right)^2 - \frac{\Omega^2}{4}} . \quad (5)$$

Considering the relations obtained, one realizes that the most important parameter determining the emission temporal behavior is the ratio $R = \gamma_{cav}/\Omega$ [Fig. 1(c)]. Well-developed Rabi oscillations are observed for $R \ll 1$. In this case, the out-of-cavity emission rate oscillates accordingly, and the emission process stretches over long time periods. In the opposite limit, $R \gg 1$, there is a long non-oscillatory response, and it is found that the emission rate is reaching maximum values and the emission takes the shortest time, when $R_{opt} \cong 1.1$ (see Supplementary Material) – roughly the critical coupling condition,

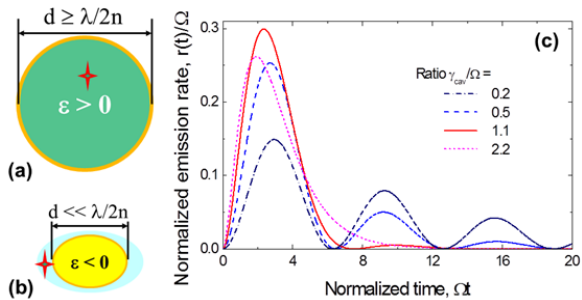

Fig. 1. (a) Dielectric cavity with a QE placed inside. (b) Plasmonic nanostructure with a QE located within a tightly confined LSP field. (c) Normalized out-of-cavity QE emission rate for different ratios between the cavity emission rate and Rabi frequency.

with the optimum cavity quality factor given by

$$Q_{opt} = \frac{\omega}{R_{opt}\Omega} \cong \frac{\pi}{1.1}\sqrt{\frac{\omega}{6\gamma_0}} = \frac{\pi}{1.1}\sqrt{\frac{\pi\nu}{3\gamma_0}} \quad . \quad (6)$$

Here, we introduced the frequency $\nu = \omega/2\pi$ in order to facilitate example calculations. The above relation is practically the same as that derived within the weak-coupling approximation [Eq. (3)], and we conclude that the time required for a photon to leave the *optimum* cavity is $T \cong 2\pi/\Omega$, and thus the maximum rate of photons is

$$B_{\max} \cong \frac{\Omega}{2\pi} = \frac{\sqrt{6\gamma_0\omega}}{2\pi^2} = \frac{\sqrt{3\pi\gamma_0\nu}}{\pi^2} \quad . \quad (7)$$

The above relation represents the *fundamental diffraction-determined limit* for the photon rate out of dielectric cavities.

Considering, for example, a QE with the lifetime of 10 ns, i.e., with $\gamma_0 = 10^8$ s$^{-1}$, and the SE being centered at the wavelength of 1 $\mu$m, i.e., $\nu \cong 3 \cdot 10^{14}$ s$^{-1}$, one obtains from Eq. (6) that the optimum quality factor (for the diffraction-limited cavity) should be ~ 5100, which would ensure the maximum rate of single photons of ~ 54 GHz [Eq. (7)]. Note that, for larger cavities, both these values should be proportionally modified: the cavity (optimum) quality factor should be larger resulting consequently in a lower (out-of-cavity) emission rate [see Eq. (3)]. From the viewpoint of Rabi oscillations, larger cavities imply weaker vacuum fields and thus smaller Rabi frequencies, which in turn require smaller optimum cavity emission rates and larger quality factors [see Eq. (6)]. At any rate, this level of cavity quality factors has already been realized and even exceeded bringing QE-cavity systems in the strong-coupling regime [5,6].

Let us now consider the QE coupling to a generic LSP sustained by a plasmonic nanostructure [7]. The *fundamental* issue with this configuration is related to the fact that the LSP quality factor is relatively low and principally limited (in the electrostatic approximation [8]) by the electron collision frequency $\gamma_m \sim 10^{14}$ s$^{-1}$ in metals, when adopting the Drude model for describing the metal dielectric function [9]. One should also take into account the radiation channel of the LSP dissipation (characterized by the emission rate $\gamma_{rad}$). When a QE interacts efficiently with an LSP field, i.e., when the QE is sufficiently close to the corresponding plasmonic nanostructure, photons are emitted primarily via the LSP radiation [10]. The SE rate of the QE-LSP system can therefore be written in the weak-coupling approximation as follows:

$$\gamma_{SE} = F\frac{\gamma_{rad}}{\gamma_m + \gamma_{rad}}\gamma_0 = \frac{3}{4\pi^2}\left(\frac{\lambda}{n}\right)^3\frac{\omega\gamma_{rad}}{V_{LSP}(\gamma_m + \gamma_{rad})^2}\gamma_0 . \quad (8)$$

Here, the LSP volume $V_{LSP}$ should be understood as an effective volume occupied by the LSP field, whose calculation is, in general, a complicated issue due to energy dissipation [11], but whose value (for strongly confined modes) is typically of the same order of magnitude as the nanostructure volume itself. Also the Purcell factor should be used with care when considering plasmonic nanostructures [12].

The LSP emission rate can be estimated by the considering the LSP being due to an electrical dipole resonance [13], with free electrons in metal oscillating (without dissipation) and generating the corresponding dipole moment (see Supplementary Material). Introducing the effective nanostructure volume:

$$V_{eff} = \left|\int \mathbf{E}(\mathbf{r})d^3r\right|^2 \Big/ \int E^2(\mathbf{r})d^3r , \quad (9)$$

we can link the dipole magnitude and the LSP mode energy and find the emission rate using the classical formula for the radiating dipole:

$$\gamma_{rad} = \frac{4}{3}\pi^2\omega\frac{V_{eff}}{\lambda\lambda_p^2} , \quad (10)$$

with $\lambda_p = 2\pi c/\omega_p$ and the validity domain being $V_{eff} \ll \lambda\lambda_p^2$.

Relating the LSP mode volume $V_{LSP}$ associated with the Purcell factor [Eq. (8)] and the effective nanostructure volume $V_{eff}$, which for a spherical nanoparticle is simply equal to the particle volume, is a challenging issue that can hardly be dealt with in a simple and general way. Since these volumes are of the same order of magnitude for highly confined modes that we are interested in, we assume hereafter that $V_{LSP} \cong V_{eff}$. Combining Eqs. (8) and (10), we obtain a key result – the *fundamental loss-determined limit* for LSP-enhanced photon rates:

$$\gamma_{SE} = \left(\frac{\lambda}{n}\right)^3 \frac{\omega^2}{\lambda \lambda_p^2 (\gamma_m + \gamma_{rad})^2} \gamma_0 \leq \frac{1}{n^3} \cdot \frac{\omega_p^2}{\gamma_m^2} \gamma_0 = \gamma_{SE}^{\max}, \quad (11)$$

with the validity domain ($\gamma_{SE} \ll \gamma_m$) imposed by the weak-coupling approximation. It should be understood that the upper limit cannot be physically reached as it requires the zero LSP volume and placing the QE on the metal surface (of a metal nanostructure sustaining the corresponding LSP). Considering a realistic case when the LSP radiative decay equals its absorption decay, $\gamma_{rad} = \gamma_m$, one obtains a (realistic) limit decay rate:

$$\gamma_{SE}^{rl} = 0.25 \frac{1}{n^3} \cdot \frac{\omega_p^2}{\gamma_m^2} \gamma_0 . \quad (12)$$

Considering the same QE as above, $n = 1$ and a silver(gold)-based LSP nanostructure characterized by $\omega_p \cong 9.6(8.55)\,\text{eV}$ and $\gamma_m \cong 22.8(18.4)\,\text{meV} \cong 5.5(4.4) \cdot 10^{12}\,\text{s}^{-1}$ [15], one obtains from Eq. (12) the maximum SE rate $\gamma_{SE}^{rl} \cong 4.43(5.4) \cdot 10^{12}\,\text{s}^{-1} \sim \gamma_m$, setting thus the maximum rate of single-photons at ~ 4.4(5.4) THz, which is *two orders of magnitude larger* than that obtained above for dielectric cavities. One can also use Eq. (10) to deduce that this SE rate requires the LSP volume corresponding to a ~ 10-nm-radius spherical nanoparticle. Recently, ultrafast ($\gamma_{SE}^{-1} \cong 13 \cdot 10^{-12}\,\text{s}$) single-photon SE was demonstrated with quantum dots coupled to gap-plasmon based nanocavities [16], and large SE enhancements in metal nanostructures (found using the antenna *RLC*-circuit approach) were suggested for improving the performance of light-emitting diodes [17]. It should further be noted that the difference in the limits obtained for these two classes would, for a given metal, *increase* for QEs with shorter lifetimes radiating at longer wavelengths, since $\gamma_{SE}^{rl}/B_{\max} \propto \sqrt{\gamma_0/\omega}$ [cf. Eqs. (7) and (12)]. Finally, the estimated SE rate is seen just at the limit of the weak-coupling approximation, indicating that the strong-coupling regime ($\gamma_{SE} \gg \gamma_0$) is within the reach for strongly confined QE-LSP configurations as indeed was very recently demonstrated [18].

Let us now turn our attention to the SE enhancement for QEs located in waveguides, starting with the dielectric case [Fig. 2(a)]. If a waveguide mode is strongly confined as, e.g., in high dielectric contrast ridge, nanowire and photonic crystal waveguides, the SE occurs mainly into the propagating waveguide modes with the rate enhancement that can be described by the Purcell factor for waveguides [19]:

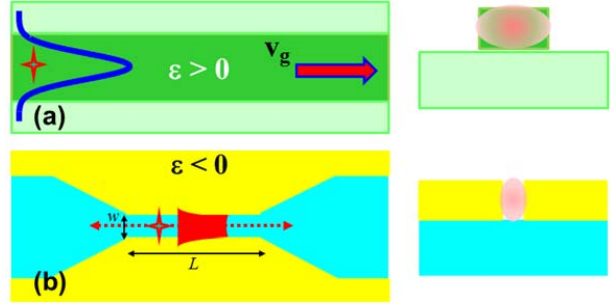

Fig. 2. Schematic configurations of (a) dielectric waveguide with a QE inside, and (b) tapered gap SPP waveguide configuration providing a strong mode confinement at the place where a QE is located. Possible realizations of the corresponding two-dimensional waveguides are illustrated to the right, depicting cross sections perpendicular to the propagation direction and the locations of the supported mode fields.

$$F_w \cong \frac{1}{\pi} \left(\frac{\lambda}{2n}\right)^2 \frac{n_g}{A_{wm}} \Rightarrow \gamma_{SE} \cong F_w \gamma_0 , \quad (13)$$

where $n_g$ is the mode group index and $A_{wm}$ is the modes size, i.e., the mode cross sectional area whose definition is a somewhat complicated issue [19]. Under the condition of diffraction-limited performance one obtains the upper limit for the Purcell factor:

$$A_{wm} \geq \left(\frac{\lambda}{2n}\right)^2 \Rightarrow F_w \leq \frac{n_g}{\pi} \quad . \quad (14)$$

The only possibility to significantly increase the SE rate into diffraction-limited waveguide modes is therefore to make use of slow-down effects that can conveniently be realized with photonic crystal waveguides (ensuring also tight mode confinement) near the band edge [20]. The *fundamental* issue with this configuration is related to the fact that the slow-down effect is of a very narrow bandwidth causing also a drastic increase in the propagation losses, so that even an optimistic estimate would be $n_g \leq 100$ [20,21]. Consequently, this implies that the Purcell factor is at best limited by 30 with the maximum rate of photons estimated (for the same QE) to be < 3 GHz.

We now turn our attention to the plasmonic waveguides supporting propagation of surface plasmon-polariton (SPP) modes, laterally confined far beyond the diffraction limit [19,22]. The *fundamental* issue with this configuration is related to the fact that the propagation loss in SPP-based waveguides increases drastically for strongly confined modes. This problem, known since the very inception of research in plasmonics, can be mitigated by coupling a strongly confined SPP waveguide to a low-loss (dielectric) waveguide before the SPP energy is dissipated in the metal [23]. Typically, one would first adiabatically taper out a very narrow lossy SPP waveguide to a relatively wider and lower loss SPP waveguide as has been demonstrated in [25] an subsequently couple to a dielectric, e.g., Si-based, waveguide [Fig. 2(b)] as has already been successfully and efficiently realized in gap SPP waveguides [24]. Taking into account the propagation loss incurred in the narrowest part of a plasmonic waveguide while neglecting the power loss elsewhere (i.e., to absorption and radiation out of the waveguide as well as during propagation in adiabatic tapers and coupling to lossless photonic waveguides), the SE rate can be written by modifying Eq. (13) as follows:

$$\gamma_{SE} = F_w \gamma_0 \exp\left(-L/L_{SPP}\right) \quad , \quad (15)$$

where $L$ is the length of the most narrow part of a plasmonic waveguide [Fig. 2(b)] with the SPP mode being characterized by the wavelength $\lambda_{SPP}$ and the propagation length $L_{SPP}$.

Let us consider a tapered configuration supporting gap SPP (GSP) modes [25], leaving its coupling to a wider GSP waveguide [Fig. 2(b)] and further to a dielectric waveguide out of analysis. Another similar configuration is a V-groove, or indeed any trench waveguide, with the width $w$ being in this case an averaged trench width. Using the limiting case of very small gap width ($w \ll \lambda$) for approximating the GSP wavelength [22] one can estimate the corresponding Purcell factor with the help of Eq. (13) as follows:

$$F_w \cong \left\| \lambda_{GSP} \cong \frac{\pi |\varepsilon_m^|| w}{\varepsilon_d}; A_{wm} \cong \frac{w\lambda_{GSP}}{2} \right\| \cong \frac{\varepsilon_d}{\pi^3} \cdot \frac{\lambda_p}{\lambda}\left(\frac{\lambda_p}{w}\right)^3 . \quad (16)$$

Here $\varepsilon_m^|$ is the real part of the metal permittivity, $\varepsilon_d = n^2$, and the mode area is approximated by $A_{GSP} \cong 0.5\lambda_{GSP}w$. The $1/w^3$ scaling in [Eq. (16)] is similar to $1/R^3$ scaling found for metal nanowires [23], signifying the fact that very large Purcell factors can be achieved with plasmonic waveguides.

Considering the same system parameters as in the above case of QE-LSP coupling and the GSP configuration with a challenging but reasonable gap of 4 nm (e.g., a 0.9-nm-wide gap was realized in the recent experiments [18]), one obtains $\lambda_p \cong 130(145)$ nm and, consequently, $F_w \cong 144(223)$. The latter values are significantly larger than the best estimate for photonic crystal waveguides, and the $1/w^3$ scaling indicates that even much larger values of the Purcell factor are within the reach. The presence of the exponential loss factor in the expression for the enhanced SE rate [Eq. (15)] emphasizes the importance of a proper choice of the narrow gap length $L$. It seems reasonable to suggest that this length should be close to the mode wavelength [26]: $L \sim \lambda_{SPP}$, so that the role of the loss factor can be neglected (for silver and gold, $L_{GSP}/\lambda_{GSP} \cong \omega/(4\pi\gamma_m) > 1$), at least in the present estimations.

Plasmonics offers unique possibilities for the manipulation of light at the nanoscale resulting in extreme light concentration and giant local field enhancements, phenomena that can advantageously be exploited in many fundamental and applied disciplines, including quantum optics. The field of *quantum plasmonics* is still relatively new [7], and its case is yet to be presented and tried, given the inevitable dissipation found in any plasmonic configuration [13]. In this Letter, we attempted to analyze "pros" and "contras" for a particular problem in quantum optics, viz., the realization of efficient and bright single-photon sources that would enable the generation of single photons with high repetition rates. We have considered QE coupling to dielectric cavities (waveguides) and localized (propagating) SPPs assessing fundamental factors that limit the achievable SE rates in these configurations. It has been found that the latter allows one to obtain the SE rate larger by almost two orders of magnitude than the former one. It is worthwhile to note (see Supplementary Material) that the optimized metal structure with today's lossy metals offer SE rate enhancements that are just a few times below the theoretical maximum attainable in the hypothetical [14] limit of lossless plasmonic structures. It is our view that QE enhancement, where the rate rather than overall external efficiency (as in the case of LED) of emission is the ultimate measure of performance, is one of the few application niches where plasmonics can shine despite the inherent metal loss. We believe that the present analysis will also be of great help when looking for new plasmonic configurations and materials to be exploited in quantum plasmonics.

**Funding.** European Research Council (ERC) (341054); ARO Grant W911NF-15-1-0629.

See Supplement 1 for supporting content.

## REFERENCES

1. J. L. O'Brien, Science **318,** 1567 (2007).
2. M. Pelton, Nat. Photon. **9**, 427 (2015).
3. L. Novotny and B. Hecht, *Principles of Nano-Optics*, 2$^{nd}$ edn. (Cambridge Univ. Press, 2012).
4. M. Fox, *Quantum Optics: An Introduction* (Oxford Univ. Press, 2006).
5. J. P. Reithmaier, G. Sęk, A. Löffler, C. Hofmann, S. Kuhn, S. Reitzenstein, L. V. Keldysh, V. D. Kulakovskii, T. L. Reinecke, and A. Forchel, Nature **432**, 197 (2004).
6. T. Yoshie, A. Scherer, J. Hendrickson, G. Khitrova, H. M. Gibbs, G. Rupper, C. Ell, O. B. Shchekin, and D. G. Deppe, Nature **432**, 200 (2004).
7. M. S. Tame, K. R. McEnery, Ş. K. Özdemir, J. Lee, S. A. Maier, and M. S. Kim, Nat. Photon. **9**, 329 (2013).
8. F. Wang and Y. R. Shen, Phys. Rev. Lett. **97**, 206806 (2006).
9. S. A. Maier, *Plasmonics: Fundamentals and Applications* (Springer, New York, 2007).
10. A. Trügler and U. Hohenester, Phys. Rev. B **77**, 115403 (2008).
11. C. Sauvan, J. P. Hugonin, I. S. Maksymov, and P. Lalanne, Phys. Rev. Lett. **110**, 237401 (2013).
12. A. F. Koenderink, Opt. Lett. **35**, 4208 (2010).
13. G. Sun, J. B. Khurgin, and R. A. Soref, J. Opt. Soc. Am. B **25**, 1748 (2008).
14. J. B. Khurgin, Nat. Nanotech. **10**, 2 (2015).
15. M. G. Blaber, M. D. Arnold, and M. J. Ford, J. Phys. Chem. C **113**, 3041 (2009).
16. T. B. Hoang, G. M. Akselrod, and M. H. Mikkelsen, Nano Lett. **16**, 270 (2016).
17. K. L. Tsakmakidis, R. W. Boyd, E. Yablonovitch, and X. Zhang, Opt. Express **24**, 17916 (2016).
18. R. Chikkaraddy, B. de Nijs, F. Benz, S. J. Barrow, O. A. Scherman, E. Rosta, A. Demetriadou, P. Fox, O. Hess, and J. J. Baumberg, Nature **535**, 127 (2016).
19. R. F. Oulton, G. Bartal, D. F. P. Pile, and X. Zhang, New J. Phys. **10**, 105018 (2008).
20. T. Baba, Nat. Photonics **2**, 465 (2008).
21. H. C. Nguyen, S. Hashimoto, M. Shinkawa, and T. Baba, Opt. Express **20**, 22465 (2012).
22. Z. Han and S. I. Bozhevolnyi, Rep. Prog. Phys. **76**, 016402 (2013).
23. D. E. Chang, A. S. Sørensen, P. R. Hemmer, and M. D. Lukin, Phys. Rev. Lett. **97**, 053002 (2006).
24. J. Tian, S. Yu, W. Yan, and M. Qiu, Appl. Phys. Lett. **95**, 013504 (2009).
25. D. K. Gramotnev and S. I. Bozhevolnyi, Nat. Photon. **8**, 13-22 (2014).
26. D. F. P. Pile and D. K. Gramotnev, Appl. Phys. Lett. **89**, 041111 (2009).
27. E. Bermúdez-Ureña, C. Gonzalez-Ballestero, M. Geiselmann, R. Marty, I. P. Radko, T. Holmgaard, Y. Alaverdyan, E. Moreno, F. J. García-Vidal, S. I. Bozhevolnyi, and R. Quidant, Nat. Commun. **6**, 7883 (2015).

## SUPPLEMENTARY

This document provides supplementary information to "Fundamental limitations in spontaneous emission rate of single-photon sources," *Optica* volume, first page (year). It details the derivations of the fundamental limits for the spontaneous emission rate out of dielectric cavities within both weak and strong coupling approximations as well as that of the emission rate for localized surface plasmons due to resonant electrical dipoles sustained by plasmonic nanostructures under the condition of of the nanostructure volume being sufficiently small. Furthermore, the hypothetical case of lossless plasmonic structures is analyzed from the viewpoint of SE rate enhancement limitations.

For a non-absorbing cavity, which is characterized by the quality factor $Q$ and volume $V$ and contains a properly located and oriented quantum emitter (QE), and assuming that the cavity is in resonance with the QE radiative transition at the wavelength $\lambda$, the Purcell factor is given by [4]:

$$F = \frac{\gamma_{SE}}{\gamma_0} = \frac{6}{\pi^2}\left(\frac{\lambda}{2n}\right)^3 \frac{Q}{V} \quad, \tag{S1}$$

determining thereby the ratio between the modified $\gamma_{SE}$ and free-space $\gamma_0$ spontaneous emission (SE) rates. Considering further dielectric, i.e., diffraction limited and lossless, cavities one can express the Purcell factor as follows:

$$F = \left\| V \geq \left(\frac{\lambda}{2n}\right)^3 ; \ V\left(\frac{2n}{\lambda}\right)^3 = \Gamma \geq 1 \right\| = \frac{6}{\pi^2}\frac{Q}{\Gamma} \quad, \tag{S2}$$

introducing the normalized cavity volume $\Gamma$. As argued in the main text, the *optimum* (from the viewpoint of maximizing the out-of-the SE rate) coupling achieved when the SE rate matches the cavity emission rate $\gamma_{cav} = \omega/Q$, where $\omega$ is the cavity resonance frequency:

$$F\gamma_0 = \gamma_{cav} \Rightarrow \frac{6}{\pi^2}\frac{Q_{opt}}{\Gamma}\gamma_0 = \frac{\omega}{Q_{opt}} \quad. \tag{S3}$$

The above expression results straightforwardly in the optimum cavity quality factor and, as a consequence [Eq. (S2)], the optimum Purcell factor along with the maximum SE rate:

$$Q_{opt} = \pi\sqrt{\frac{\Gamma\omega}{6\gamma_0}} \Rightarrow \gamma_{SE}^{\max} = F_{opt}\gamma_0 = \frac{1}{\pi}\sqrt{\frac{6\omega\gamma_0}{\Gamma}} \quad. \tag{S4}$$

Taking into account that $\Gamma \geq 1$ results immediately in Eq. (3) given in the main text.

The above condition was obtained by making use of the Purcell factor, i.e., within the weak-coupling approximation. We demonstrate below that a similar relation can be found using a more rigorous approach that considers vacuum Rabi oscillations. Introducing the cavity emission into coupled equations describing vacuum Rabi oscillations one obtains:

$$\frac{dx}{dt} = j\frac{\Omega}{2}y \quad \text{and} \quad \frac{dy}{dt} = -\frac{\gamma_{cav}}{2}y + j\frac{\Omega}{2}x \quad. \tag{S5}$$

Here $|x(t)|^2$ is the QE excited state population (i.e., the probability of finding the QE in the excited state), $|y(t)|^2$ is the photon number inside the cavity (i.e., the probability of finding the photon inside the cavity), and $\Omega = \sqrt{6\gamma_0\omega}/\pi$ is the vacuum Rabi frequency in the diffraction-limited cavity, i.e., with the volume $V = (\lambda/2n)^3$. Using the initial conditions, $x(0) = 1$ and $y(0) = 0$, the solution can readily be found:

$$x(t) = \frac{s_2 e^{s_1 t} - s_1 e^{s_2 t}}{s_2 - s_1}, \quad y(t) = \frac{2}{j\Omega} \cdot \frac{s_1 s_2}{s_2 - s_1}\left[e^{s_1 t} - e^{s_2 t}\right], \tag{S6}$$

with parameters $s_{1,2}$ defined as

$$s_{1,2} = -\frac{\gamma_{cav}}{4} \mp \sqrt{\left(\frac{\gamma_{cav}}{4}\right)^2 - \frac{\Omega^2}{4}} \quad. \tag{S7}$$

The photon flux out of the cavity is given by $r(t) = \gamma_{cav}|y(t)|^2$, leading straightaway to Eq. (4) given in the main text. Integrating the photon flux in time, one finds the number of emitted photons, i.e., the probability of finding an emitted photon (that is left the cavity) in our case of a single excited QE inside a cavity. It is seen that for cavities with high quality factors and low cavity emission rates ($\gamma_{cav} \ll \Omega$), one finds well-developed Rabi oscillations with a QE and cavity exchanging the energy so that it takes a long time for a photon to leave the cavity [Fig. S1(a)]. For cavities with low quality factors and high cavity emission rates ($\gamma_{cav} \gg \Omega$), the population decay is almost purely exponential with a photon leaving a cavity very soon after it has been emitted into it by a QE, but the emission-out-of-cavity time is again much larger than the period of Rabi oscillations [Fig. S1(c)]. The emission is found to take the shortest time for $R_{opt} \cong 1.1$, when the probability of emitting (out of the cavity) a photon reaches 0.98 after the time period $T \cong 2\pi/\Omega$ [Fig. S1(b)].

The localized surface plasmon (LSP) emission rate that causes the LSP dissipation by radiation can be estimated by considering the LSP being due to an electric dipole resonance [12], with free electrons in metal moving (without resistance/damping) in the electrical field $\mathbf{E}(\mathbf{r})\cos(\omega t)$ so that the displacement of each electron in a metal nanostructure supporting the LSP is given by

$$\mathbf{r}(t) = \frac{e\mathbf{E}(\mathbf{r})}{m\omega^2}\cos(\omega t) \quad, \tag{S8}$$

where $e$ and $m$ are the electron charge and mass, $\omega$ is the frequency of the driving electromagnetic field. The energy of a single electron inside the metal nanostructure reads

$$u = \frac{1}{2}m\omega^2 r^2 = \frac{1}{2}\frac{e^2 E^2(\mathbf{r})}{m\omega^2} \quad. \tag{S9}$$

The total energy of the LSP mode can be evaluated via the kinetic energy of moving electrons and expressed therefore as follows:

$$U = \frac{1}{2}\frac{e^2 n}{m\omega^2}\int E^2(\mathbf{r})d^3r = \frac{1}{2}\varepsilon_0\frac{\omega_p^2}{\omega^2}\int E^2(\mathbf{r})d^3r \quad, \tag{S10}$$

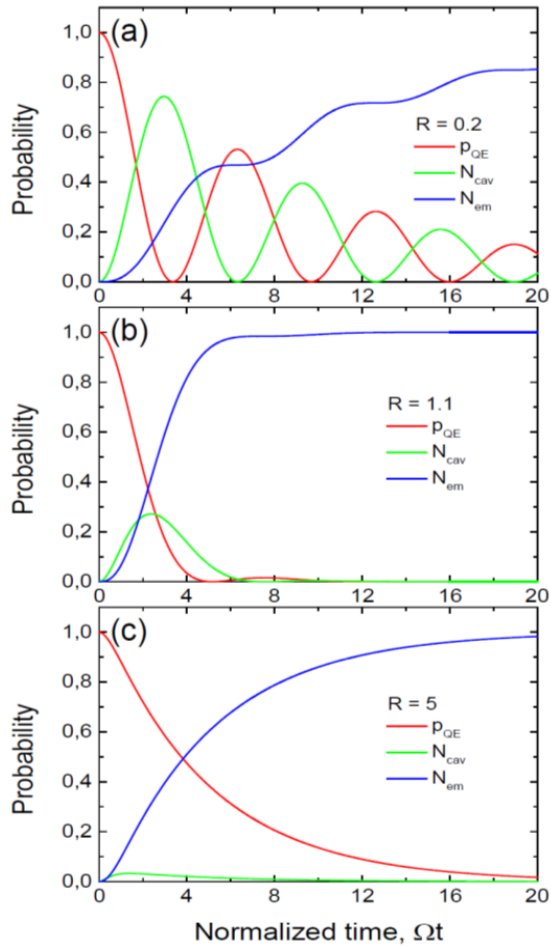

Fig. S1. Probabilities of finding a QE in the excited state ($p_{QE}$), a photon inside a cavity ($N_{cav}$) and an emitted photon ($N_{em}$) that is left the cavity for different ratios $R$ between the cavity emission rate $\gamma_{cav}$ and vacuum Rabi frequency $\Omega$: $R = \gamma_{cav}/\Omega$.

where $n$ is the electron density and $\omega_p$ is the plasma frequency, with the integration being carried out over the nanostructure volume. At the same time, the total dipole moment associated with this LSP mode is given by

$$\mathbf{p}(t) = -en\int \mathbf{r}(t)d^3r = -\frac{e^2 n}{m\omega^2}\int \mathbf{E}(\mathbf{r})d^3r \cos(\omega t) \;, \quad \textbf{(S11)}$$

so that the dipole magnitude can be expressed as follows:

$$p_0 = |\mathbf{p}_0| = \frac{e^2 n}{m\omega^2}\left|\int E^2(\mathbf{r})d^3r\right| = \varepsilon_0 \frac{\omega_p^2}{\omega^2}\left|\int E^2(\mathbf{r})d^3r\right| \;. \quad \textbf{(S12)}$$

The above relations [Eqs. (S10) and (S12)] allow us to establish the connection between the LSP energy and the dipole magnitude:

$$p_0^2 = 2\varepsilon_0 \frac{\omega_p^2}{\omega^2}\frac{\left|\int \mathbf{E}(\boldsymbol{r})d^3r\right|^2}{\int E^2(\boldsymbol{r})d^3r}U = 2\varepsilon_0 \frac{\omega_p^2}{\omega^2}V_{em}U \;, \quad \textbf{(S13)}$$

where we have introduced the effective LSP emission volume:

$$V_{em} = \left|\int \mathbf{E}(\mathbf{r})d^3r\right|^2 \Big/ \int E^2(\mathbf{r})d^3r \;. \quad \textbf{(S14)}$$

Note that for a simple spherical nanoparticle the field is uniform inside the particle, and the emission volume is thus equal to the volume of the nanoparticle itself. Finally, we make use of the classical formula for the power of the dipole radiating into the free space in order to find the LSP emission rate $\gamma_{rad}$:

$$P_{rad} = \frac{p_0^2 \omega^4}{12\pi\varepsilon_0 c^3} = \frac{\omega^2 \omega_p^2 V_{em}}{6\pi c^3}U = -\gamma_{rad}U \;, \quad \textbf{(S15)}$$

which results in the final expression [see also Eq. (10)]:

$$\gamma_{rad} = \frac{\omega^2 \omega_p^2 V_{em}}{6\pi c^3} = \frac{4}{3}\pi^2 \omega \frac{V_{em}}{\lambda \lambda_p^2} \;. \quad \textbf{(S16)}$$

The above derivation disregards higher order multipoles in the LSP excitation and thus can only be used when $V_{em} \ll \lambda \lambda_p^2$.

Considering fact that the SE rate in QE-LSP systems is fundamentally limited by the electron collision frequency in metals, one might surmise that it is actually advantageous to have *large* absorption, which should of course be appropriately balanced by the plasma frequency [Eq. (12)]:

$$\gamma_{SE}^{rl} = 0.25\frac{1}{n^3}\cdot\frac{\omega_p^2}{\gamma_m^2}\gamma_0 < \gamma_m \; \text{ or } \; \omega_p < 2n\gamma_m\sqrt{n\gamma_m/\gamma_0} \;. \quad \textbf{(S17)}$$

This conclusion might seem surprising, since any dissipation should and does decrease the photon emission in any form. In order to reveal the influence of loss in this particular configuration of quantum plasmonics, we analyze below the *hypothetical* case of lossless plasmonic nanostructures. Using the results [Eqs. (1) and (10)] from the main document, one finds the LSP quality factor:

$$Q = \frac{\omega}{\gamma_{rad}} = \frac{3}{4\pi^2}\frac{\lambda \lambda_p^2}{V} \;. \quad \textbf{(S18)}$$

This brings us to the expression for the Purcell factor:

$$F = \left(\frac{3}{4\pi^2}\right)^2 \left(\frac{\lambda}{n}\right)^3 \frac{\lambda \lambda_p^2}{V^2} \;. \quad \textbf{(S19)}$$

This situation is somewhat similar to that discussed for dielectric micro-cavities: better quality factors imply smaller radiation rates, and the optimum condition is as follows:

$$F\gamma_0 = \left(\frac{3}{4\pi^2}\right)^2 \left(\frac{\lambda}{n}\right)^3 \frac{\lambda \lambda_p^2}{V_{opt}^2} = \gamma_{rad} = \frac{4}{3}\pi^2 \omega \frac{V_{opt}}{\lambda \lambda_p^2} \quad . \qquad \textbf{(S20)}$$

The optimum volume can then be written down as follows:

$$V_{opt} = \frac{3}{4\pi^2 n}\left(\frac{\gamma_0}{\omega}\right)^{1/3} \lambda^3 \left(\frac{\lambda_p}{\lambda}\right)^{4/3} \quad . \qquad \textbf{(S21)}$$

Substituting Eq. (S21) into Eq. (10) results in the expression for the *fundamental radiation-determined limit* for lossless LSP-enhanced photon rates:

$$\gamma_{SE}^{\max} = n^{-1}\left(\gamma_0 \omega^2\right)^{1/3} \left(\frac{\lambda}{\lambda_p}\right)^{2/3} = n^{-1}\left(\gamma_0 \omega_p^2\right)^{1/3} \quad . \qquad \textbf{(S22)}$$

Considering the same QE as before and the typical plasma frequency $\omega_p = 1.45 \times 10^{16}\,\text{s}^{-1}$, we obtain from the above relation: $\gamma_{SE}^{\max} \cong 18\,\text{THz}$ and the volume corresponding to a 14-nm-radius spherical nanoparticle. It is thus seen that the lossless plasmonic nanostructures are *definitely better* (by 4 times) in terms of SE rate enhancement than the realistic ones, as one could have expected [see Eq. (12)]. It is though still surprising that realistic plasmonic losses are not that bad, since their reduction from real numbers all the way down to zero would improve the situation *only* by a factor of 4 - both limits are staying within the same order of magnitude.